\documentclass[11pt]{article}
\usepackage{jhep}

\usepackage[utf8]{inputenc}
\usepackage{amsmath,amssymb}
\usepackage[separate-uncertainty=true]{siunitx}
\usepackage{graphicx}
\usepackage[usenames,dvipsnames]{xcolor}
\usepackage[caption=false]{subfig}
\usepackage{hyperref}
\usepackage[capitalize]{cleveref}
\usepackage{booktabs}
\usepackage{tabularx}
\usepackage{xspace}
\usepackage[normalem]{ulem}

\usepackage{graphicx}
\usepackage{flushend}
\usepackage{soul}
\usepackage{amsmath}
\usepackage{amssymb}
\usepackage{array}
\usepackage{tabularx}
\usepackage{snapshot}

\allowdisplaybreaks

\AtBeginDocument{
\heavyrulewidth=.08em
\lightrulewidth=.05em
\cmidrulewidth=.03em
\belowrulesep=.65ex
\belowbottomsep=0pt
\aboverulesep=.4ex
\abovetopsep=0pt
\cmidrulesep=\doublerulesep
\cmidrulekern=.5em
\defaultaddspace=.5em

\newcolumntype{C}{>{\centering\arraybackslash}X}
\newcolumntype{b}{C}
\newcolumntype{s}{>{\hsize=.6\hsize}C}
\newcolumntype{R}{>{\raggedleft\arraybackslash}X}
}

\newcommand{\fb}{{\ensuremath\rm fb}}
\newcommand{\pb}{{\ensuremath\rm pb}}

\newcommand{\GeV}{{\ensuremath\rm GeV}}
\newcommand{\MeV}{{\ensuremath\rm MeV}}
\newcommand{\TeV}{{\ensuremath\rm TeV}}

\newcommand{\vareps}{\varepsilon}

\newcommand{\htb}[1]{{\color{blue} #1}}

\newcommand{\lb}{\left(}
\newcommand{\rb}{\right)}
\newcommand{\lam}{\lambda}
\newcommand{\eqdot}{\,.}
\newcommand{\eqcomma}{\,,}

\begin{document}
\title{\boldmath Triple Higgs Boson Production at the Large Hadron Collider with Two Real Singlet Scalars}

\bibliographystyle{JHEP} 

\emailAdd{apapaefs@cern.ch}
\emailAdd{trobens@irb.hr}
\emailAdd{gtx@physik.uni-siegen.de}

\date{Revised version: \today}

\author[a, b]{Andreas Papaefstathiou,}

\author[c]{Tania Robens,}

\author[d]{Gilberto Tetlalmatzi-Xolocotzi,}

\affiliation[a]{Higgs Centre for Theoretical Physics, University of Edinburgh, Peter Guthrie Tait Road, Edinburgh EH9 3FD, UK.}
  
\affiliation[b]{Department of Physics, Kennesaw State University, Kennesaw, GA 30144, USA.}

\affiliation[c]{Ruder Boskovic  Institute,  Bijenicka  cesta  54,  10000  Zagreb, Croatia.}

\affiliation[d]{Theoretische Physik 1, Naturwissenschaftlich-Technische Fakult\"{a}t, Universit\"{a}t Siegen,Walter-Flex-Strasse 3, D-57068 Siegen, Germany.}

\abstract{We investigate the production of three Higgs bosons in the Two Real Singlet extension of the Standard Model, where the scalar sector is augmented by two additional real scalar fields which are singlets under the Standard Model gauge group. The model contains three neutral CP-even scalars, allowing for resonant production and asymmetric decay chains. We focus on the signature $p p\,\rightarrow\,h_3\,\rightarrow\,h_1\,h_2\,\rightarrow\,h_1\,h_1\,h_1$, where we identify $h_3$ as the heaviest scalar state, $h_2$ as the second heaviest and the lightest, $h_1$, as the Standard Model-like Higgs boson discovered by the Large Hadron Collider experiments. The dominant final state occurs when all three Higgs bosons decay to bottom-anti-bottom quark pairs, $h_1\,\rightarrow\,b\,\bar{b}$, leading to 6 $b$-jets. Taking into account all current theoretical and experimental constraints, we determine the discovery prospects for this channel in future runs of the Large Hadron Collider, as well as in the high-luminosity phase.}

\preprint{ 
	
	\vspace{-0.45cm}
	
	\begin{flushright}
	     SI-HEP-2020-34\\ RBI-ThPhys-2020-53  \\
	     PH3-21-006\\
	\end{flushright}
}

\maketitle

\section{Introduction}
With the discovery of a particle which complies with the expected properties of the Higgs boson of the Standard Model (SM) by the CERN Large Hadron Collider (LHC) experiments in 2012 \cite{Aad:2012tfa,Chatrchyan:2012ufa}, particle physics has entered an exciting new era. Although current experimental results agree rather well with the predictions of the SM, both experimental and theoretical uncertainties allow for new phenomena that may be observable either at current or future colliders. In the present article we focus on the particularly interesting possibility of models that extend the scalar sector of the SM by additional scalar fields that transform as singlets under the SM gauge group. Such models may provide solutions to a multitude of fundamental open questions: they could contain viable candidates of dark matter or enable mechanisms that could explain the observed cosmic matter-anti-matter asymmetry (see, e.g.~\cite{Gil:2012ya,Carena:2012np,Profumo:2014opa,Kozaczuk:2014kva,Jiang:2015cwa, Vaskonen:2016yiu,Kotwal:2016tex,Dorsch:2016nrg,Chiang:2017nmu,deVries:2017ncy,Beniwal:2017eik,Beniwal:2018hyi,Bruggisser:2018mrt,Athron:2019teq,Kainulainen:2019kyp,Ellis:2019flb,Ramsey-Musolf:2019lsf,Bian:2019kmg,Xie:2020bkl,Papaefstathiou:2020iag}). They are also very rich in terms of their collider phenomenology, introducing new physical scalar states that can participate in cascade decays.

In this work, we concentrate on the triple production of scalar final states resulting from the asymmetric decay chain:
\begin{equation}\label{eq:chain}
p p \,\rightarrow\, h_3\,\rightarrow\,h_2\,h_1\,\rightarrow\,h_1\,h_1
\,h_1,
\end{equation}
where $h_{1,2,3}$ are the physical scalar states of a model with an extended scalar sector. We require that one of these scalars, specifically the $h_1$ boson, is identified with the 125~\GeV{} SM Higgs particle, including agreement with all current measurements. The other scalars, however, can lie in any mass range, as long as all theoretical and experimental constraints are satisfied. As we are interested in the discovery potential of colliders that probe the \TeV{} scale, we choose to consider scenarios with masses $\lesssim\,1\,\TeV$. 

In order to allow for the decay chain (\ref{eq:chain}), and assuming CP conservation, the new physics model under consideration needs to contain at least three CP-even scalar states. One of the simplest ways to realise this is through models that extend the SM scalar sector by \textit{two} additional singlet fields. The two real\footnote{Models with two real singlets or one complex singlet field are equivalent, given that potential additional symmetries are correctly translated, see, e.g.~\cite{Barger:2008jx,Robens:2019kga}.} singlet extension that contains three unstable physical scalars has been widely investigated in the literature, see, e.g.~\cite{Barger:2008jx,AlexanderNunneley:2010nw,Coimbra:2013qq,Ahriche:2013vqa,Costa:2014qga,Costa:2015llh,Ferreira:2016tcu,Chang:2016lfq,Muhlleitner:2017dkd,Dawson:2017jja,Robens:2019kga,Barducci:2019xkq,Englert:2020ntw,aali:2020tgr,Atkinson:2020uos} for recent discussions. 

The LHC experimental collaborations have already largely scrutinised models which allow for several scalar particles in the final state, including searches for processes with symmetric di-scalar production via resonances, $p\,p\,\rightarrow\,h_2\,\rightarrow\,h_1\,h_1$, where either $h_1$ or $h_2$ take the role of the 125 \GeV~ SM-like scalar \cite{Aaboud:2016xco,Aaboud:2016oyb,Sirunyan:2017djm,Sirunyan:2017guj,Aaboud:2018knk,Aaboud:2018ftw,Aaboud:2018ewm,Aaboud:2018sfw,Aaboud:2018zhh,Aaboud:2018ksn,Aaboud:2018gmx,Aaboud:2018iil,Aaboud:2018esj,Sirunyan:2018mbx,Sirunyan:2018iwt,Sirunyan:2018zkk,Sirunyan:2018two,Sirunyan:2018mgs,Sirunyan:2018mot,Aad:2019uzh,Aad:2020kub,Aad:2020rtv,Sirunyan:2020eum,Sirunyan:2020qcq}. Furthermore, in~\cite{Aaboud:2018ksn} the ATLAS collaboration also interpreted their results for the above production and decay chain for pure beyond-the-SM (BSM) scalars, i.e.\ neither $h_{1}$ nor $h_2$ assume the role of the SM Higgs boson. For models with extended scalar sectors, however, triple couplings between different mass states, $\lam_{h_i h_j h_k}$, can best be probed at leading order in resonant production modes such as the decay chain (\ref{eq:chain}). Such states have e.g. been discussed in \cite{King:2014xwa,Costa:2015llh,Ellwanger:2017skc,Baum:2018zhf,Baum:2019uzg,Englert:2020ntw,Atkinson:2020uos}, but currently no experimental results for such searches are available.

While the investigation of the process $p p \rightarrow h_1\,h_2$ with decays into SM-like final states is an important quest as such,\footnote{For representative benchmark points for such scenarios, see e.g.~\cite{Robens:2019kga}.} here we plan to focus on the specific case where $h_2\,\rightarrow\,h_1\,h_1$, leading to triple scalar final states as indicated above. In the SM, the production cross section for the triple Higgs boson final state is the lowest-order process to include the quartic Higgs self-coupling. At the LHC's nominal  centre-of-mass energy, 14~TeV, the corresponding cross section in the SM is diminishingly small, $\sim\,0.1\,\fb$~\cite{Maltoni:2014eza,deFlorian:2016sit}, rising up to a cross section of $\sim\,5.6\,\fb$ at a 100 \TeV{} proton-proton collider~\cite{deFlorian:2019app}. While the quartic self-coupling in the SM can also be indirectly constrained \cite{Bizon:2018syu,Borowka:2018pxx,Liu:2018peg,DiMicco:2019ngk}, direct determination seems to call for future high-energy proton-proton colliders \cite{Papaefstathiou:2015paa,Chen:2015gva,Dicus:2016rpf,Contino:2016spe,Fuks:2017zkg,Papaefstathiou:2019ofh} or a possible muon collider \cite{Maltoni:2018ttu,Chiesa:2020awd}.

As discussed above, the simplest realisation that achieves (\ref{eq:chain}) are models that extend the SM by two additional real scalar fields, which are singlets under the SM gauge group. We consider here a specific version, the ``Two Real Singlet Model'' (TRSM)~\cite{Robens:2019kga}, where in addition two $\mathbb{Z}_2$ symmetries are imposed, leading to a reduction of the available number of degrees of freedom. In the TRSM, the gluon-fusion $pp \rightarrow h_1 h_1 h_1$ cross section is enhanced via the resonant production of $h_3$ and can reach up to $140\,\fb$ at the LHC.\footnote{This prediction results from a factorised approach, where the $h_3$ production cross section has been obtained by rescaling NNLO+NNLL production cross sections for a SM-like Higgs boson at the respective mass~\cite{deFlorian:2016spz}.} While direct searches for an SM-only triple Higgs boson production are not very promising at current centre-of-mass energies, we will show that several benchmark points of the TRSM are within a $2-4\,\sigma$ significance range with an integrated luminosity of $300\,\fb^{-1}$, reaching up to $\sim\,5\,\sigma$ for selected points, and can reach up to $\sim 16\, \sigma$ for the full high-luminosity LHC (HL-LHC) nominal dataset of $3000\,\fb^{-1}$.

This article is organised as follows: in section~\ref{sec:pheno}, we briefly review the model under consideration as well as the specific benchmark plane that our study focusses on. In section~\ref{sec:const}, we discuss current theoretical and experimental constraints. The event generation, cross sections and selection analysis are discussed in section~\ref{sec:gen}. We present the results of our analysis in section~\ref{sec:res}. There we also present projections for the sensitivity of the full HL-LHC run for searches of heavy scalars within the TRSM into di-boson final states. Our summary and conclusions can be found in section~\ref{sec:summ}.

\section{The Two Real Singlet Extension of the Standard Model}\label{sec:pheno}

\subsection{Extending the Standard Model by Real Singlet Scalar Fields}

The scalar potential of the SM can be extended by an additional sector of 
scalar fields that transform as singlets under the SM gauge group, leading to

\begin{eqnarray}
V(\Phi, \phi_i)&=&V_{\rm singlets}(\Phi, \phi_i) + V_{\rm SM}(\Phi)\;,
\end{eqnarray}	

with the most general renormalizable expression for $V_{\rm singlets}(\Phi, \phi_i)$ given by

\begin{eqnarray}
V_{\rm singlets}(\Phi, \phi_i) &=& a_i \phi_i + m_{ij}\phi_i \phi_j + T_{ijk}\phi_i\phi_j\phi_k
+ \lambda_{ijkl}\phi_i\phi_j\phi_k\phi_l  \nonumber\\
&&+ T_{i H H}\phi_i(\Phi^{\dagger}\Phi)
+ \lambda_{ij H H}\phi_i \phi_j(\Phi^{\dagger}\Phi)\;.
\end{eqnarray}

In this work we focus on the TRSM \cite{Robens:2019kga}, which introduces two extra real scalar fields $S$ and $X$. The number of free parameters is constrained by imposing the following discrete $\mathbb{Z}_2$  symmetries:

\begin{eqnarray}
\mathbb{Z}^S_2 &:& S\rightarrow -S\;,\quad X \rightarrow X\;, \nonumber\\
\mathbb{Z}^X_2 &:& X\rightarrow -X\;,\quad S \rightarrow S\;,
\label{eq:symm}
\end{eqnarray}

and where all SM particles transform evenly under both symmetries.

The application of the discrete symmetries of eq.~(\ref{eq:symm}) reduces the scalar potential for two real singlet fields to:

\begin{eqnarray}\label{eq:reduced}
V(\Phi, X, S) &=&\mu^2_{\Phi}\Phi^{\dagger}\Phi +\lambda_{\Phi}\Bigl(\Phi^{\dagger}\Phi\Bigl)^2
+\mu^2_S S^2 +\lambda_S S^4 + \mu^2_X X^2  + \lambda_X X^4  \nonumber\\
&&  +\lambda_{\Phi S}\Phi^{\dagger}\Phi S^2 +
\lambda_{\Phi X} \Phi^{\dagger}\Phi X^2 + \lambda_{S X} S^2 X^2\;,
\end{eqnarray}

which is characterised by nine real couplings $\mu_{\Phi}$, $\lambda_{\Phi}$,
$\mu_S$, $\lambda_S$, $\mu_X$, $\lambda_X$, $\lambda_{\Phi S}$,  $\lambda_{\Phi X}$,  $\lambda_{X S}$. All fields are assumed to acquire a vacuum expectation value (vev). The physical gauge-eigenstates $\phi_{h,S,X}$ then follow from expanding around these according to:

\begin{equation}
    \Phi = \begin{pmatrix} 0\\\frac{\phi_h + v}{\sqrt{2}}\end{pmatrix}\;,
    S = \frac{\phi_S + v_S}{\sqrt{2}}\;, \quad
    X = \frac{\phi_X + v_X}{\sqrt{2}}\;.
    \label{eq:fields}
\end{equation}

In this study we consider the broken phase in which $v_S, v_X\neq 0$
and $v = v_{\rm SM} \simeq 246$~GeV.
Then, the discrete symmetries $\mathbb{Z}^S_2$ and $\mathbb{Z}^X_2$ are spontaneously 
broken and the scalars $\phi_h$, $\phi_S$, $\phi_X$ mix into the physical 
states $h_1$, $h_2$ and $h_3$ according to

\begin{equation}
    \begin{pmatrix}
        h_1 \\h_2\\h_3
    \end{pmatrix} = R \begin{pmatrix}
        \phi_h \\\phi_S\\\phi_X
    \end{pmatrix}\;,
\end{equation}

with the rotation matrix $R$ given by

\begin{equation}\label{eq:Rmat}
    R = \begin{pmatrix}
        c_1 c_2             & -s_1 c_2             & -s_2     \\
        s_1 c_3-c_1 s_2 s_3 & c_1 c_3+ s_1 s_2 s_3 & -c_2 s_3 \\
        c_1 s_2 c_3+s_1 s_3 & c_1 s_3-s_1 s_2 c_3  & c_2 c_3
    \end{pmatrix}\;.
\end{equation}

To simplify our discussion we have used the following notation when writing $R$ in eq.~(\ref{eq:Rmat}):

\begin{eqnarray}
s_1\equiv\sin\theta_{hS}\;,&\quad& s_2\equiv\sin\theta_{hX}\;,\quad s_3\equiv\sin\theta_{SX}\;,
\nonumber\\	
c_1\equiv\cos\theta_{hS}\;,&\quad& c_2\equiv\cos\theta_{hX}\;,\quad c_3\equiv\cos\theta_{SX}\;,
\end{eqnarray}

with 

\begin{eqnarray}
-\frac{\pi}{2} < \theta_{hS}\;, \theta_{hX}\;, \theta_{SX} < \frac{\pi}{2}\;.
\end{eqnarray}

Using the same notation as in \cite{Robens:2019kga}, the entries of the first row in 
the matrix $R$ are denoted as $\kappa_i\equiv R_{i 1}$ for $i=1,2,3$. 

In principle, any of the three scalars can take the role of the SM-like Higgs boson resonance discovered by the LHC experiments, as long as the other parameters are set such that all experimental constraints are fulfilled. Here, however, we will focus on the scenario where the state $h_1$ is identified with the SM-like Higgs boson, and $h_2$ and  $h_3$ are two new physical \textit{heavier} scalars obeying the mass hierarchy

\begin{eqnarray}
M_1 \leq M_2\leq M_3\;.
\end{eqnarray}

As previously described, there are $9$ real parameters characterising the TRSM. However,
the identification of $h_1$ as the SM Higgs boson fixes 

\begin{center}
\begin{eqnarray}
M_1&\backsimeq& 125\,\rm{GeV},\nonumber\\
v&\backsimeq&246\,\rm{GeV}.
\end{eqnarray}
\end{center}

This leaves us with $7$ independent parameters, which we chose as

\begin{eqnarray}
M_2\;, M_3\;, \theta_{h S}\;, \theta_{h X}\;, \theta_{S X}\;, v_S, v_X\;.	
\label{eq:freeparam}
\end{eqnarray}

In this model all couplings for the mass eigenstates $h_i$ to SM particles are inherited from the SM-like Higgs doublet through the rotation from the gauge to the mass eigenstates, such that $g_i\,\equiv\,\kappa_i\,g^\text{SM}$. For example, in a factorised approach, this leads to predictions for production cross sections of the form
\begin{equation}
\sigma\lb p p \rightarrow h_i \rb\,=\,\kappa_i^2\,\sigma^\text{SM}\lb p p \rightarrow h_\text{SM} \rb\,\lb M_i\rb, 
\end{equation}
 where $\sigma^\text{SM}\lb M_i \rb$ denotes the production cross section of an SM-like Higgs boson of mass $M_i$.

Furthermore, the total width of the $h_i$ scalars ($i=1,2,3$) is given by:
\begin{equation}\label{eq:totwidth}
\Gamma_{h_i} = \kappa_{i}^2 ~ \Gamma^{\mathrm{SM}} (M_i) + \sum_{j,k \neq i} \Gamma_{h_i \rightarrow h_j h_k},
\end{equation}
where $\Gamma^{\mathrm{SM}} (M_i)$ corresponds to the width of a scalar boson of mass $M_i$ possessing the same decay modes as a SM Higgs of mass $M_i$. The branching ratios corresponding to $h_i \rightarrow xx$, for $x \neq h_j$ ($j\neq i)$ are then given by:
\begin{equation}
\mathrm{BR}(h_i \rightarrow x x) = \kappa_{i}^2 \frac{\Gamma_{xx}^{\mathrm{SM}} (M_i)}{\Gamma_{h_i}} \;,
\end{equation}
where $\Gamma_{xx}^{\mathrm{SM}} (M_i)$ corresponds to the SM-like partial decay width of a scalar boson of mass $M_i$ for the final state $xx$.
The scalar-to-scalar branching ratios are equivalently obtained via 
\begin{equation}
\mathrm{BR}(h_i \rightarrow h_j h_k) =  \frac{\Gamma_{h_i\,\rightarrow\,h_j\,h_l}}{\Gamma_{h_i}}.
\end{equation}

\subsection{Benchmark Scenario}

As discussed in \cite{Robens:2019kga}, depending on the values 
that the free parameters of eq.~(\ref{eq:freeparam}) assume, different realisations 
of the TRSM are possible, yielding a rich phenomenology at colliders. Here we concentrate on the ``Benchmark Plane 3'' {\bf{(BP3)}} addressed in~\cite{Robens:2019kga}, which was carefully tailored to allow for a large region in the $(M_2, M_3)$ plane which obeys all current theoretical and experimental constraints, while at the same time allowing for a large $h_1 h_1 h_1$ decay rate.\footnote{Note that, in addition, this rate depends on the mixing angles and additional vevs, which are fixed in \textbf{BP3}.} \textbf{BP3} is characterised by the numerical values of the parameters shown in table \ref{tab:BP3}.\footnote{Note that we actually set $M_1\,=\,125\,\GeV$ in the analysis performed throughout this work.}

\begin{table}
\centering
\begin{tabular}{cc}
Parameter & Value \\	
\hline
$M_1$ & $125.09~\rm{GeV}$\\
$M_2$ & $[125,~500]~\rm{GeV}$\\
$M_3$ & $[255,~650]~\rm{GeV}$\\
$\theta_{hS}$ & $-0.129$\\
$\theta_{hX}$ & $0.226$ \\
$\theta_{SX}$ & $-0.899$\\
$v_S$ & $140~\rm{GeV}$\\
$v_X$ & $100~\rm{GeV}$\\ \hline
$\kappa_1$ & $0.966$\\
$\kappa_2$ & $0.094$\\
$\kappa_3$ & $0.239$
\end{tabular}	
\caption{\label{tab:BP3} The numerical values for the independent parameter values of eq.~(\ref{eq:freeparam}) that characterise \textbf{BP3}. The Higgs doublet vev, $v$, is fixed to 246~GeV. The $\kappa_i$ values correspond to the rescaling parameters of the SM-like couplings for the respective scalars and are derived quantities.}
\end{table}

\section{Constraints and Allowed Regions}\label{sec:const}
 Constraints on the TRSM have been discussed in detail in~\cite{Robens:2019kga}, and we essentially follow that description in this work. In particular, we include constraints from perturbative unitarity, the requirement that the potential is bounded from below and agreement with electroweak precision observables. Results from null searches at colliders for the additional resonances as well as agreement with the current signal strength measurements, have been tested using the \texttt{HiggsBounds}~\cite{Bechtle:2008jh,Bechtle:2011sb,Bechtle:2013gu,Bechtle:2013wla,Bechtle:2015pma,Bechtle:2020pkv} and \texttt{HiggsSignals}~\cite{Bechtle:2013xfa, Stal:2013hwa, Bechtle:2014ewa,Bechtle:2020uwn} packages. We additionally made use of the \texttt{ScannerS}~\cite{Coimbra:2013qq,Costa:2015llh,Muhlleitner:2020wwk} code to cross-check several of the constraints discussed in this section. In the rest of this section, we describe the constraints in further detail.

\subsection{Theory Constraints}

We can derive constraints on the values that the masses $M_2$ and $M_3$
can assume by considering the perturbative unitarity of the $2\,\rightarrow\,2$ scalar scattering matrix in the TRSM. Moreover, we impose an upper limit $|\mathcal{M}_i|\,\leq\,8\,\pi$ on the eigenvalues $\mathcal{M}_i$ of the scattering matrix $\mathcal{M}$.

These limits can be written in terms of the coupling constants as\footnote{For further details on the derivation of the limits in terms of the coupling constants, see e.g.\ the discussion in \cite{Wittbrodt:2019bsu}.}

\begin{eqnarray}
|\lambda_{\Phi}|< 4\pi \;,\nonumber\\
|\lambda_{\Phi S}|\;,|\lambda_{\Phi X}|\;, |\lambda_{S X}|< 8\pi \;,\nonumber\\
|a_1|\;, |a_2|\;, |a_3|<16\pi \;,
\end{eqnarray}

where $a_{1,2,3}$  correspond to the roots of the following polynomial:

\begin{eqnarray}
P(x)&=& x^3 +x^2(-12 \lambda_{\Phi}-6 \lambda_S- 6\lambda_X)
	+ x \left[72 \lambda_{\Phi} (\lambda_S + \lambda_X)- 
	4(\lambda^2_{\Phi S}+\lambda^2_{\Phi X})\right.\nonumber\\
	&&\left. +36 \lambda_S \lambda_X - \lambda^2_{SX}\right]
	+ 12\lambda_{\Phi} \lambda^2_{S X} + 24 \lambda^2_{\Phi S}\lambda_X
	+24 \lambda^2_{\Phi X} \lambda_S -8 \lambda_{\Phi S} \lambda_{\Phi X} \lambda_{SX}
	-432\lambda_{\Phi} \lambda_S \lambda_X\;. \nonumber\\
&&
\end{eqnarray}

The potential of eq.~(\ref{eq:reduced}) additionally needs to be bounded from below. This requirement was implemented in the scan discussed in \cite{Robens:2019kga} using the conditions derived in \cite{Kannike:2012pe,Kannike:2016fmd}, which we list here for completeness
\begin{equation}
    \begin{aligned}
        \lambda_\Phi,\lambda_S,\lambda_X                                                                                               & > 0\eqcomma  \\
        \overline{\lambda}_{\Phi S} \equiv \lambda_{\Phi S} + 2 \sqrt{\lambda_\Phi\lambda_S}                                           & > 0 \eqcomma \\
        \overline{\lambda}_{\Phi X} \equiv \lambda_{\Phi X} + 2 \sqrt{\lambda_\Phi\lambda_X}                                           & > 0 \eqcomma \\
        \overline{\lambda}_{SX} \equiv \lambda_{SX} + 2 \sqrt{\lambda_S\lambda_X}                                                      & > 0 \eqcomma \\
        \sqrt{\lambda_S}\lambda_{\Phi X}
        + \sqrt{\lambda_X}\lambda_{\Phi S}
        + \sqrt{\lambda_\Phi}\lambda_{SX}
        + \sqrt{\lambda_\Phi\lambda_S\lambda_X} + \sqrt{\overline{\lambda}_{\Phi S}\overline{\lambda}_{\Phi X}\overline{\lambda}_{SX}} & > 0\eqdot
    \end{aligned} 
\end{equation}

These constraints are especially important for masses in the region $M_2\,\lesssim\,140\,\GeV,\,M_3\,\in\,\left[500, 650\right]$ \GeV. However, the most dominant theoretical bound in this plane stems from perturbative unitarity.

\subsection{Electroweak Precision Constraints}
In the benchmark plane discussed here, constraints from electroweak precision observables have been imposed using the \texttt{ScannerS} interface, which calculates the oblique parameters $S,T,U$ \cite{Altarelli:1990zd,Peskin:1990zt,Peskin:1991sw,Maksymyk:1993zm} from expressions in \cite{Grimus:2007if,Grimus:2008nb} and compares them to the most recent fit results of the GFitter collaboration \cite{Haller:2018nnx}, including all correlations. 
\subsection{Collider Constraints}

To apply current constraints we employ the \texttt{HiggsBounds} (v5.9.0) and \texttt{HiggsSignals} (v2.5.1) packages.  \texttt{HiggsBounds} takes a selection of Higgs sector predictions for any model as input and then uses the experimental topological cross-section limits from Higgs boson searches at LEP, the Tevatron and the LHC to determine if this parameter point has been excluded at 95\% C.L.. \texttt{HiggsSignals} performs a statistical test of the Higgs sector predictions of arbitrary models with the measurements of Higgs boson signal rates and masses from the Tevatron and the LHC. \texttt{HiggsBounds} returns a boolean corresponding to whether the Higgs sector passes the constraints at 95\% C.L. (true) or not (false). \texttt{HiggsSignals} returns a probability value ($p$-value) corresponding to the goodness-of-fit of the Higgs sector over several SM-like ``peak'' observables.  The code contains searches up to the full LHC Run II luminosity, and we refer the reader to the documentation of the code for details \cite{hbgitlab}.

For \textbf{BP3}, we found that searches for $h_{2,3}\,\rightarrow\,V\,V$ from 2016 LHC Run II data \cite{Aaboud:2017rel,Sirunyan:2018qlb,Aaboud:2018bun} constrain some parts of the parameter space, in agreement with the results presented in \cite{Robens:2019kga}.

\section{Event Simulation and Analysis}\label{sec:gen}

\subsection{Monte Carlo Event Generation}\label{sec:mc}
All the parton-level events used in the phenomenological analysis of the present article have been generated via the Monte Carlo (MC) event generator \texttt{MadGraph5\_aMC@NLO} (v2.7.3)~\cite{Alwall:2014hca, Hirschi:2015iia}. The TRSM signal MC samples were produced via a custom modification of the \texttt{loop\_sm} model to incorporate the additional scalar particles and their interactions with the SM particles. This yields a leading-order description of the signal, including the full top and bottom quark mass dependence and all interference effects between the contributing Feynman diagrams. The production of the samples for the background process, i.e.\ the final state that originates from the QCD production of $(b\bar{b}) (b\bar{b}) (b\bar{b})$ constitutes the most challenging aspect of the event generation. Note that within the SM this entails the evaluation of 6762 Feynman diagrams. To address this challenge we heavily parallelised the event generation via the ``gridpack'' option provided by \texttt{MadGraph5\_aMC@NLO}. 

QCD parton showering, hadronization and underlying event simulation were all performed within the general-purpose MC event generator \texttt{HERWIG} (v7.2.1)~\cite{Bahr:2008pv, Gieseke:2011na, Arnold:2012fq, Bellm:2013hwb, Bellm:2015jjp, Bellm:2017bvx, Bellm:2019zci}. Events were subsequently analysed via the \texttt{HwSim} module~\cite{hwsim} for \texttt{HERWIG} which saves events in a \texttt{ROOT} compressed file format~\cite{Brun:1997pa}, with jets clustered using \texttt{FastJet} (v3.3.2)~\cite{Cacciari:2011ma}. The anti-$k_T$ algorithm~\cite{Cacciari:2008gp} with a radius parameter $R=0.4$ was used to cluster jets. A detailed study of pile-up effects arising from secondary proton-proton interactions is beyond the scope of the present phenomenological study and will need to be addressed in a full experimental study that will include in conjunction a detailed description of detector effects. For a recent a discussion on the issue of pile-up mitigation and corrections, we would like to point out the reader to the detailed studies of Ref.~\cite{Soyez_2019}, which demonstrate the degree of the effects on jet resolution and suggest approaches in the form of advanced techniques to improve on this.

To capture the detector effects, we only consider particles with transverse momentum $p_T > 100$~MeV as being detectable. We do not consider any smearing of momenta coming from detector mis-measurements. Similarly, we do not take possible mis-identification of light or charm jets as $b$-jets into account. These assumptions are not expected to have a dramatic impact on the conclusions of the present study and we anticipate that a full experimental analysis will assess their effects in detail. Throughout this work, we assume a $b$-jet tagging efficiency of 0.7, which lies on the conservative side of 13~TeV ATLAS and CMS performance~\cite{Aad:2015ydr,CMS-DP-2017-012,ATLAS-CONF-2018-045} and was also adopted in the studies presented in \cite{Cepeda:2019klc}. We have elected to consider a constant $b$-tagging efficiency with transverse momentum and pseudo-rapidity of the jets. This is justified since, for example, by examining Fig. 6 of Ref.~\cite{ATLAS-CONF-2018-045}, where the $b$-tagging efficiency appears to be relatively flat for both observables, and in particular above $p_T \sim 30$~GeV, with $\mathcal{O}(10\%)$ uncertainty, which is precisely where we impose a cut on the $b$-jets in our analysis.

\subsection{Cross Sections}

\begin{center}
\begin{figure}
\begin{center}
 \includegraphics[width=0.65\columnwidth]{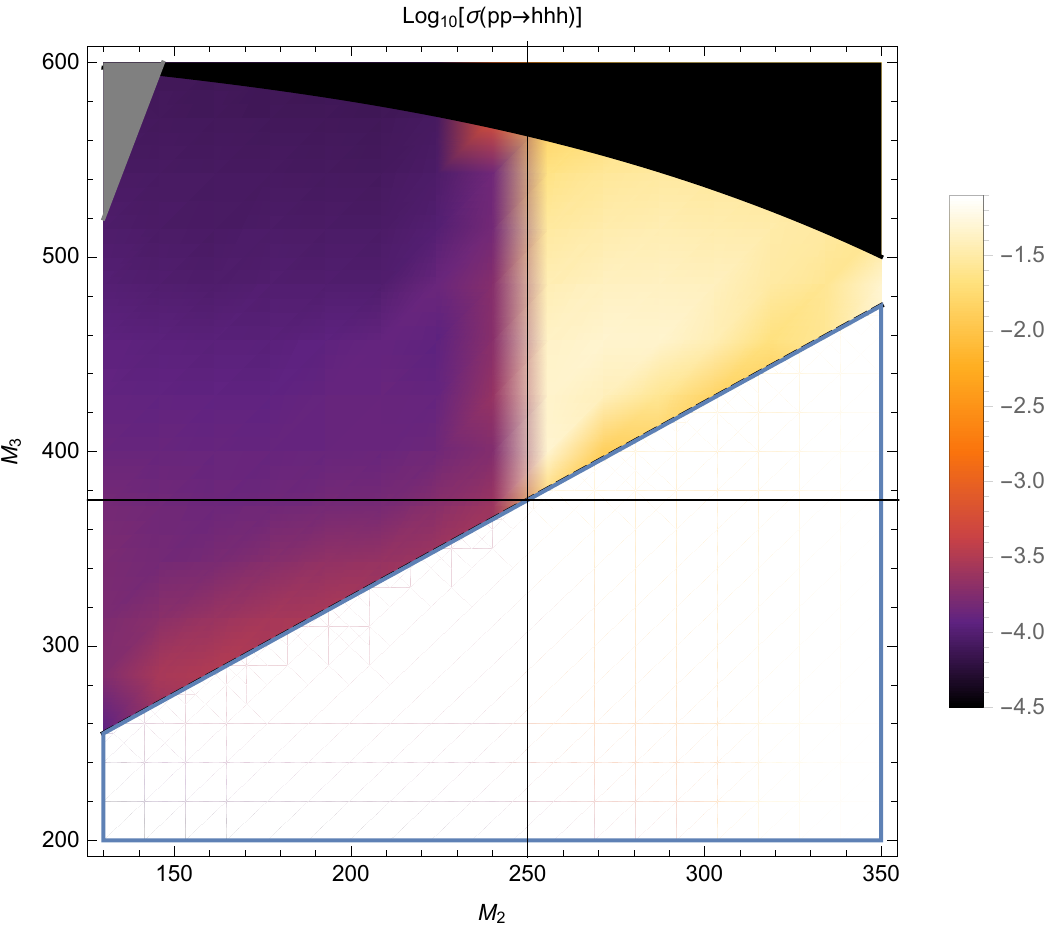}
\caption{The total leading-order gluon-fusion production cross sections for the $p\,p\,\rightarrow\,h_1\,h_1\,h_1$ process at a 14 \TeV~ LHC. No cuts have been imposed. We also show the region excluded by constraints coming from perturbative unitarity in the dark upper part and boundedness from below in the gray wedge. In the allowed region, the leading-order predictions reach cross-section values of up to $\sim 50 \,\fb$. }
\label{fig:crosssections_m2m3}
\end{center}
\end{figure}
\end{center}

We present the production cross sections for the $ pp \rightarrow h_1 h_1 h_1$ final state over \textbf{BP3} in fig.~\ref{fig:crosssections_m2m3}, where in addition  bounds from perturbative unitarity and the requirement for the potential to be bounded from below are shown. The cross sections displayed in this plot have been obtained following the leading-order MC description of section~\ref{sec:mc}, which includes all gluon-fusion-initiated contributions as well as interference effects (e.g.\ the box diagrams $gg \rightarrow h_1 h_1 h_1$, $gg \rightarrow h_2 \rightarrow h_1 h_1 h_1$ or $gg \rightarrow h1 \rightarrow h_1 h_1 h_1$). Note that, for points where indeed the $h_3$ resonant production contributes dominantly, one could additionally apply a K-factor to account for missing higher-order contributions, e.g.\ with respect to the NNLO+NNLL corrected predictions for production cross sections of an SM-like scalar with mass $M_3$~\cite{deFlorian:2016spz}. For our selected benchmark points within \textbf{BP3}, specified below, we found that these K-factors for gluon-gluon induced $h_3$ production are $\sim\,2.5$.\footnote{For parameter points where $h_3$ production dominates, the total cross section is in addition sensitive to the total width of $h_3$ and follows the scaling predicted by the narrow width approximation, i.e. $\sigma_{p p \rightarrow\,h_1 h_1 h_1}\,\sim\,\Gamma_3^{-1}$. Therefore, percent-level differences in the width can induce similar changes in the final result.} Furthermore, for all of our benchmark points we found that $\sim\;93-99\%$ of the cross section stems from the decay chain specified in eq.~(\ref{eq:chain}).  
\begin{table}
\begin{center}

\begin{tabular*}{0.7\textwidth}{@{\extracolsep{\fill}}cccc@{}}
Label&$(M_2, M_3)$& $\sigma(pp\rightarrow h_1 h_1 h_1)$ &
$\sigma(pp\rightarrow 3 b \bar{b})$\\
&$\rm{[GeV]}$ & $\rm{[fb]}$  & $\rm{[fb]}$  \\
\hline
 \textbf{A}&$(255, 504)$ & $32.40$ & $6.40$\\
\textbf{B}&$(263, 455)$ & $50.36$ & $9.95$\\
\textbf{C}&$(287, 502)$ & $39.61$ & $7.82$\\
\textbf{D}&$(290, 454)$ & $49.00$ & $9.68$\\
\textbf{E}&$(320, 503)$ & $35.88$& $7.09$\\
\textbf{F}&$(264, 504)$ & $37.67$ & $7.44$\\
\textbf{G}&$(280, 455)$& $51.00$ & $10.07$ \\
\textbf{H}& $(300, 475)$&$43.92$& $8.68$\\\
\textbf{I}&$(310, 500)$& $37.90$ & $7.49$ \\
\textbf{J}&$(280, 500)$& $40.26$& $7.95$\\
\end{tabular*}
\caption{The leading-order gluon-fusion production cross sections for the $pp \rightarrow h_1 h_1 h_1$ signal for different realisations of \textbf{BP3}, depending on the masses of the scalars $h_2$ and $h_3$ in the region $M_2 > 250$~GeV and $M_3 > 375$~GeV. The given combinations of masses presented are allowed by current constraints. The numbers correspond to a proton-proton centre-of-mass energy of $\sqrt{s}=14~\rm{TeV}$. The fourth column assumes mediation via the $h_1h_1h_1$ intermediate state. The statistical integration uncertainties are smaller than the accuracy shown here.}
\label{tab:XS_signal}	
\end{center}
\end{table}

For our analysis, we have selected specific benchmark points within \textbf{BP3}. The corresponding cross-section predictions for $pp \rightarrow h_1\,h_1\,h_1$ as well as $6\,b$-quark final states are given in table~\ref{tab:XS_signal}.\footnote{The widths for the three scalars have been calculated according to eq.~(\ref{eq:totwidth}), with SM-like widths from \cite{Heinemeyer:2013tqa}. We list the corresponding values in Appendix \ref{app:vals}, together with the corresponding new physics branching ratios.} Here we have taken the branching ratio of the $h_1$ to $b\bar{b}$ to be $\text{BR}_{h_1\,\rightarrow\,b\,\bar{b}}\,=\,0.5824$ \cite{deFlorian:2016spz}. The SM background amounts to a cross section of 6.38~\pb{} for the 6 $b$-quark final state from QCD-induced diagrams, including a K-factor of 2, typical for gluon-fusion processes. Additional backgrounds from electroweak processes, e.g. $Z\,b\,\bar{b}\,b\,\bar{b}$ production with $Z\,\rightarrow\,b\bar{b}$, as discussed in \cite{Papaefstathiou:2019ofh}, were found to be at least two orders of magnitude lower and have not been considered in our study. We expect that these will form a sub-dominant contribution with respect to the QCD background after the analysis cuts are imposed.\\

\subsection{Selection Analysis}\label{sec:ana}

Our analysis has been adapted from that of ref.~\cite{Papaefstathiou:2019ofh}. An event is analysed if it contains at least $6$ $b$-tagged jets\footnote{Since the Higgs bosons are produced with transverse momenta up to $\mathcal{O}(100)$~GeV, i.e.\ comparable to their mass, we do not expect the $b$-jets to frequently merge into a singlet jet and therefore we focus only on the ``resolved'' $6$ $b$-jet scenario.} with a transverse
momentum of at least $p_{Tmin, b}=25$~GeV and a pseudo-rapidity no greater than $|\eta_{b, max}|=2.5$. These initial cuts are further optimised for each of our signal samples, which are characterised by different combinations of $M_2$ and $M_3$.

We then select the $6$ $b$-tagged jets with the highest transverse momentum and form pairs in different combinations, with the aim of first reconstructing individual SM-like Higgs bosons, $h_1$, and subsequently the two scalars $h_2$ and $h_3$. To this end, we introduce two observables:

\begin{eqnarray}
\label{eq:chi4}
\chi^{2, (4)}&=&\sum_{q r \in I}\Bigl(M_{q r} - M_1 \Bigl)^2\;,
\end{eqnarray}

\begin{eqnarray}
\label{eq:chi6}
\chi^{2, (6)}&=&\sum_{q r\in J}\Bigl(M_{q r} - M_1 \Bigl)^2\;.
\end{eqnarray}

where we have defined the sets $I=\{i_1 i_2, i_3 i_4\}$ and $J=\{j_1 j_2, j_3 j_4, j_5 j_6\}$, constructed from different pairings of 4 and 6 $b$-tagged jets, respectively, and where $M_{q r}$ denotes the invariant mass of the respective pairing, $q r$. It should be understood that each jet can appear only in a single arrangement inside $I$ and $J$. The number of possible $n$ pairings given the 6 $b$-jets with the highest $p_T$ is given by $\frac{1}{n !}\,\binom{6}{2}\,\binom{4}{2}$, which translates to 45 different combinations for $I$ and 15 combinations for $J$, respectively.

We select the combinations of $b$-tagged jets entering in $I$ and $J$ based on the minimisation of the sum
 
 \begin{eqnarray}
 \chi^{2, (6)} + \chi^{2, (4)}\;.
 \end{eqnarray}
 
 The above procedure still allows for different approaches in the combination strategy, on which we briefly comment in Appendix \ref{app:combi}. We then ``identify'' candidates for the scalars $h_2$ and $h_3$ with the pairing configurations $I_{\rm min}$ and $J_{\rm min}$ which minimise $\chi^{2,(4)}$ and $\chi^{2, (6)}$ respectively, as described above. Note that this procedure does not guarantee that $I_\text{min}$ indeed reconstructs to $h_2$; in fact, we found this to be the case in about $40\%$ on average for all benchmark samples, being slightly higher than a ``blind guess'' that would lead to a probability of 1/3. Based on the invariant mass of the $b$-jet combinations entering in $I_{\rm min}$ and $J_{\rm min}$, we define two additional observables $m^{\rm inv}_{4b}$ and $m^{\rm inv}_{6b}$. We wish to stress that we do not make explicit use of the values of $M_2$ and $M_3$ for the individual samples. The fact that the masses are different is however taken into account implicitly considering that we find different selection cuts depending on the concrete signal sample during the analysis.  Our approach  is already able to deliver a good selection performance and using additional information on the assumed values for $M_2$ and $M_3$ can only improve the selection results.

Since each pairing inside $J_{\rm min}$ ``defines'' a Higgs boson candidate $h^{i}_1$, we determine the absolute differences between the invariant mass of each pairing and $M_1$, i.e.~the mass of the SM Higgs boson. Each one of these differences is sorted from minimum to maximum, $(\Delta m_{\rm min}, \Delta m_{\rm med}, \Delta m_{\rm max})$. The size of these deviations is an indicator of how accurately the individual SM Higgs bosons are reconstructed. Since $\Delta m_{\rm min}, \Delta m_{\rm med}<\Delta m_{\rm max}$, the maximum deviation from $M_1$ is precisely $\Delta m_{\rm max}$. In practice we find that our selection criteria give a distribution for $\Delta m_{\rm max}$ which peaks at about $10~\rm{GeV}$ in all the signal samples studied.

We also obtain the transverse momentum $p_T(h^i_1)$ of the $h^{i}_1$ candidate, constructed from the pairings inside $J_{\rm min}$. These transverse momenta are then ordered from hardest to softest and used as variables for signal and background discrimination. Similarly, we make use of the angular distance $\Delta R(h^i_1, h^j_1)$ between the $h_1$ candidates  $h^{i}_1$ and $h^j_1$ and additional angular cuts $\Delta R_{bb}(h^i_1) $ are enforced between the $b$-jet pairs that define each of the $h^{i}_1$ candidates.

The optimisation of the analysis is based on the sequential application of cuts on the different observables described previously, until the significance is numerically above the minimum threshold of 2. More concretely, we obtain the ``best'' selection cuts for each observable using the following order: (i) $p_{Tmin,b}$ and $|\eta_b|$, (ii) $\chi^{2, (6)}$ and $\chi^{2, (4)}$, (iii)  $m^{\rm inv}_{6b}$, (iv) $m^{\rm inv}_{4b}$.  
We finally establish the values for the selection cuts
affecting  the pairings of $b$-jets  which define $J_{\rm min}$ and $I_{\rm min}$  as follows: (v) $p_{T}(h^i_1)$, (vi) $(\Delta m_{\rm min,~med,~max})$, (vii) $\Delta R(h^i_1, h^j_1)$, (viii) $\Delta R_{bb}(h^i_1)$. 

The optimisation takes place by constructing a grid over the selection observables and exploring sequentially combinations of cuts which deliver the maximum rejection of the background while maintaining the highest acceptance for the signal. The grid is established by studying the observable distributions to deduce its limits appropriately. Specifically, we look for the maximum and minimum values that capture all the signal events. In the particular case of the invariant masses, bounds from perturbative unitarity pose an additional constraint, which allows us to define the corresponding grid. As an explicit example, the values for $p_{Tmin,b}$ and the maximum $|\eta_b|$ are obtained by calculating all the possible combinations inside the intervals $[25,~40]$~GeV and $[1.0,~2.5]$ over a $20 \times 10$ grid, respectively. Each possible cut combination is then tested over signal and background and the significance is calculated. At this stage we keep those cut combinations which deliver a significance above $1.5$. We then optimise on $\chi^{2,(6)}$ and  $\chi^{2,(4)}$ in an analogous fashion, taking as starting values for $p_{Tmin,b}$ and $|\eta_b|$ from the best pairings obtained in the first stage. At each layer of the optimisation procedure we increase the minimum threshold for the significance. In table~\ref{tab:cuts} we summarise the combination of cuts which give the best performance in our selection procedure.

\section{Results}\label{sec:res}

\subsection{Results for Triple Higgs Boson Production}

\begin{table}[t!]
\begin{tabular*}{\textwidth}{@{\extracolsep{\fill}}ccccccc@{}}
Label & $(M_2, M_3)$ & $<P_{T,b}$ & $\chi^{2, (4)}<$ & 
	$\chi^{2, (6)}<$ & $m^{\rm inv}_{4b}<$ & $m^{\rm inv}_{6b}<$ \\ 	
 & $\rm{[\GeV]}$ & $\rm{[\GeV]}$ &  $\rm[{\GeV}^2]$& $[\rm{\GeV}^2]$ & $\rm{[\GeV]}$ & 
	$\rm{[\GeV]}$ \\
\hline
\textbf{A} & $(255, 504)$& $34.0$ &  $10$ & $20$ & - & $525$\\ 
\textbf{B} & $(263, 455)$& $34.0$  & $10$& $20$ & $450$ & $470$\\
\textbf{C} & $(287, 502)$& $34.0$ & $10$ & $50$ & $454$ & $525$\\
\textbf{D} & $(290,454)$ & $27.25$  & $25$ & $20$ & $369$ & $475$\\
\textbf{E} & $(320, 503)$ & $27.25$  & $10$ & $20$ & $403$ & $525$\\ 
\textbf{F} & $(264, 504)$& $34.0$&  $10$& $40$& $454$ & $525$ \\
\textbf{G} & $(280, 455)$& $26.5$  &$25$& $20$& $335$ & $475$\\
\textbf{H} & $(300, 475)$& $26.5$ &  $15$ & $20$ & $352$ & $500$\\
\textbf{I} & $(310, 500)$& $26.5$& $15$ & $20$ & $386$ &$525$\\
\textbf{J} & $(280, 500)$& $34.0$  & $10$ & $40$ & $454$ & $525$\\
\end{tabular*}
\caption{\label{tab:cuts} The optimised selection cuts for each of the benchmark points within \textbf{BP3} shown in table~\ref{tab:XS_signal}. The cuts not shown above are common for all points, as follows: $|\eta|_b<2.35$,  $\Delta m_{\rm min,~med,~max}< [15, 14, 20]$ \GeV, $p_T(h^i_1) > [50, 50, 0]$ \GeV, $\Delta R(h^i_1, h^j_1)< 3.5$ and $\Delta R_{bb}(h_1)< 3.5$. For some of the points a $m^{\rm inv}_{4b}$ cut is not given, as this was found to not have an impact when combined with the $m^{\rm inv}_{6b}$ cut.}
\end{table}

\begin{table*}[t!]
\begin{tabular*}{\textwidth}{@{\extracolsep{\fill}}cccccccc@{}}
Label&$(M_2, M_3)$ & $\varepsilon_{\rm Sig.}$& $\rm{S}\bigl|_{300\rm{fb}^{-1}}$ & $\varepsilon_{\rm Bkg.}$ & 
$\rm{B}\bigl|_{300\rm{fb}^{-1}}$ & $\text{sig}|_{300\rm{fb}^{-1}}$ & $\text{sig}|_{3000\rm{fb}^{-1}}$\\
& [GeV] & & & & & (syst.) &(syst.)  
\\
\hline
\textbf{A} &$(255, 504)$ & $0.025$ & $14.12$  & $8.50\times 10^{-4}$ & $19.16$ & $2.92~(2.63)$&$9.23~(5.07)$\\
\textbf{B} & $(263, 455)$ & $0.019$ & $17.03$    & $3.60\times 10^{-5}$ & $ {8.12}$ & $4.78 ~(4.50)$&$15.10~(10.14)$\\
\textbf{C} & $(287, 502)$ & $0.030$ & $20.71$ & $9.13\times 10^{-5}$ & $20.60$  & $4.01~(3.56)$ & $12.68~(6.67)$\\
\textbf{D} & $(290, 454)$ & $0.044$ & $37.32$    & $1.96\times 10^{-4}$ & $44.19$& $5.02~(4.03)$&$15.86~(6.25)$\\
\textbf{E} & $(320, 503)$ & $0.051$ & $ {31.74}$    & $2.73\times 10^{-4}$ & $61.55$& $3.76~( {2.87}) $&$11.88~(4.18)$\\
\textbf{F} & $(264,504)$&$0.028$& $18.18$&$9.13\times 10^{-5}$&$20.60$&$3.56~(3.18) $&$11.27~(5.98)$\\
\textbf{G} & $(280, 455)$&$0.044$& $38.70$ &$1.96\times 10^{-4}$& $44.19$ & $5.18~(4.16)$ &$16.39~(6.45)$\\
\textbf{H} & $(300, 475)$ & $0.054$& $41.27$ & $2.95\times 10^{-4}$& $66.46$ & $4.64~(3.47)$&$ 14.68~( {4.94})$\\
\textbf{I} & $(310, 500)$& $0.063$& $41.43$& $3.97\times 10^{-4}$& $89.59$& $4.09~(2.88) $&$ {12.94~(3.87)}$\\
\textbf{J} & $(280,500)$& $0.029 $& $20.67$&$9.14\times 10^{-5}$& $20.60$&$4.00~(3.56) $&$12.65~(6.66)$\\
\end{tabular*}
\caption{ The resulting selection efficiencies, $\varepsilon_{\rm Sig.}$ and $\varepsilon_{\rm Bkg.}$, number of events, $S$ and $B$ for the signal and background, respectively, and statistical significances for the sets of cuts presented in table~\ref{tab:cuts}. A $b$-tagging efficiency of $0.7$ has been assumed. The number of signal and background events are provided at an integrated luminosity of $300~\rm{fb}^{-1}$. Results for $3000~\rm{fb}^{-1}$ are obtained via simple extrapolation. The significance is given at both values of the integrated luminosity excluding (including) systematic errors in the background according to Eq.~(\ref{eq:Sig}) (or Eq.~(\ref{eq:syst}) with $\sigma_b=0.1\times\rm{ B}$).}
\label{table:efficiencies}
\end{table*}

In table~\ref{table:efficiencies}, we list the expected number of signal and background events after the application of all cuts, as given in table~\ref{tab:cuts} for each point, where we include a $K$-factor of 2 for the background and 2.5 for the signal, as well as the corresponding selection efficiencies $\vareps$, giving the fraction of MC events that pass the cuts. We also show the predicted statistical significances at integrated luminosities of $300\,\fb^{-1}$ and $3000\,\fb^{-1}$.

Since the number of signal events $S$ and the number of background events $B$ are of the same order,  $S\,\sim\,B$, we employ the following definition of the statistical significance \cite{Cowan:2010js}

\begin{equation}
\label{eq:Sig}
\text{sig}\lb S,B\rb\,=\,\sqrt{2\,\left[\lb S+B\rb\,\ln\lb 1+S/B\rb-S\right]}.
\end{equation}

To incorporate the effects of systematic uncertainties, the significance can be estimated according to \htb{\cite{Cowan:2010js,slac12,medsig}}

\begin{equation}
\label{eq:syst}
\text{sig}\lb S,B\rb\,=\sqrt{2\Biggl(\Bigl[S+B\Bigl] \ln\Biggl[\frac{(S+B)(B + \sigma^2_B)}{B^2+ (S + B)\sigma^2_B}\Biggl]-\frac{B^2}{\sigma^2_B}\ln\Bigl[1 + \frac{\sigma^2_B S}{B(B +\sigma^2_B)}\Bigl] \Biggl)},  
\end{equation}

where $\sigma_B$ is an estimate of the systematic uncertainty on the total background contributing to this process. We will assume this to have the form $\sigma_B = \alpha B$, where we will set $\alpha = 0.1$ to represent a 10\% systematic uncertainty on the total background rates.\footnote{This is reasonable, since e.g. in \cite{Cepeda:2019klc}, $6\%$ maximal uncertainties were suggested for $b-$jet related quantities. See also \cite{uncerthl}.}


We see that already at an integrated luminosity of $300\,\fb^{-1}$ and in the absence of systematics, significances of up to $\sim$ 5$\sigma$ can be achieved for some of the chosen benchmark points. Furthermore, with the full HL-LHC integrated luminosity, all points are within discovery reach, and we obtain significances up to $\sim\,16\,\sigma$ for selected benchmark points. Once systematic errors are taken into account the  values for the significance are affected when the background is relatively large. However even for these cases, the significances for $3000~\rm{fb}^{-1}$ are nearly always above $4\sigma$.  

In general, the significance that can be achieved is correlated with the $h_1 h_1 h_1$ production cross sections given in table~\ref{tab:XS_signal}, such that points with higher cross sections have a tendency to lead to higher significances. As production cross sections are directly correlated to the mass $M_3$, in general lower masses result in higher significances. For similar masses $M_3$, the mass region $M_2\,\sim\,280-300\,\GeV$ seems to yield the best results. For parameter points with similar masses for $h_2$, on the other hand, significances can largely vary with the production cross section for $h_1 h_1 h_1$ and/or $M_3$, see e.g.\ points \textbf{B} and \textbf{F} or \textbf{G} and \textbf{J} for comparison, where in each case a smaller mass $M_3$/larger production cross section are correlated with higher significance. Note that the semi-automatised cut selection we apply, described in section~\ref{sec:ana}, optimises each event sample separately and therefore comparisons in the multivariate parameter space are not straightforward. In a more detailed investigation of points \textbf{I} and \textbf{F} we found, e.g., that a $\sim\; 6\%$ difference in a cut selection efficiency can increase the difference in significance by a factor 2. A similar behaviour can also be observed in the comparison of points \textbf{I} and \textbf{E}.

In summary, we find that in the region we consider in \textbf{BP3}, significances over $5 \,\sigma$ can already be achieved with an integrated luminosity of $300\,\fb^{-1}$ and that at the HL-LHC all points should be within discovery range. We would like to highlight  that our full optimisation strategy and our final results for the significance can be improved by using more sophisticated analysis techniques such as  machine-learning  multi-variable classifiers. However, in this work we chose not to no pursue such a strategy, since we have demonstrated that it is possible to reach a meaningful threshold for the significance by solely employing an iterative selection procedure.

\subsection{Other Channels at the HL-LHC}

\begin{center}
\begin{figure}
\begin{center}
  \includegraphics[width=0.8\columnwidth]{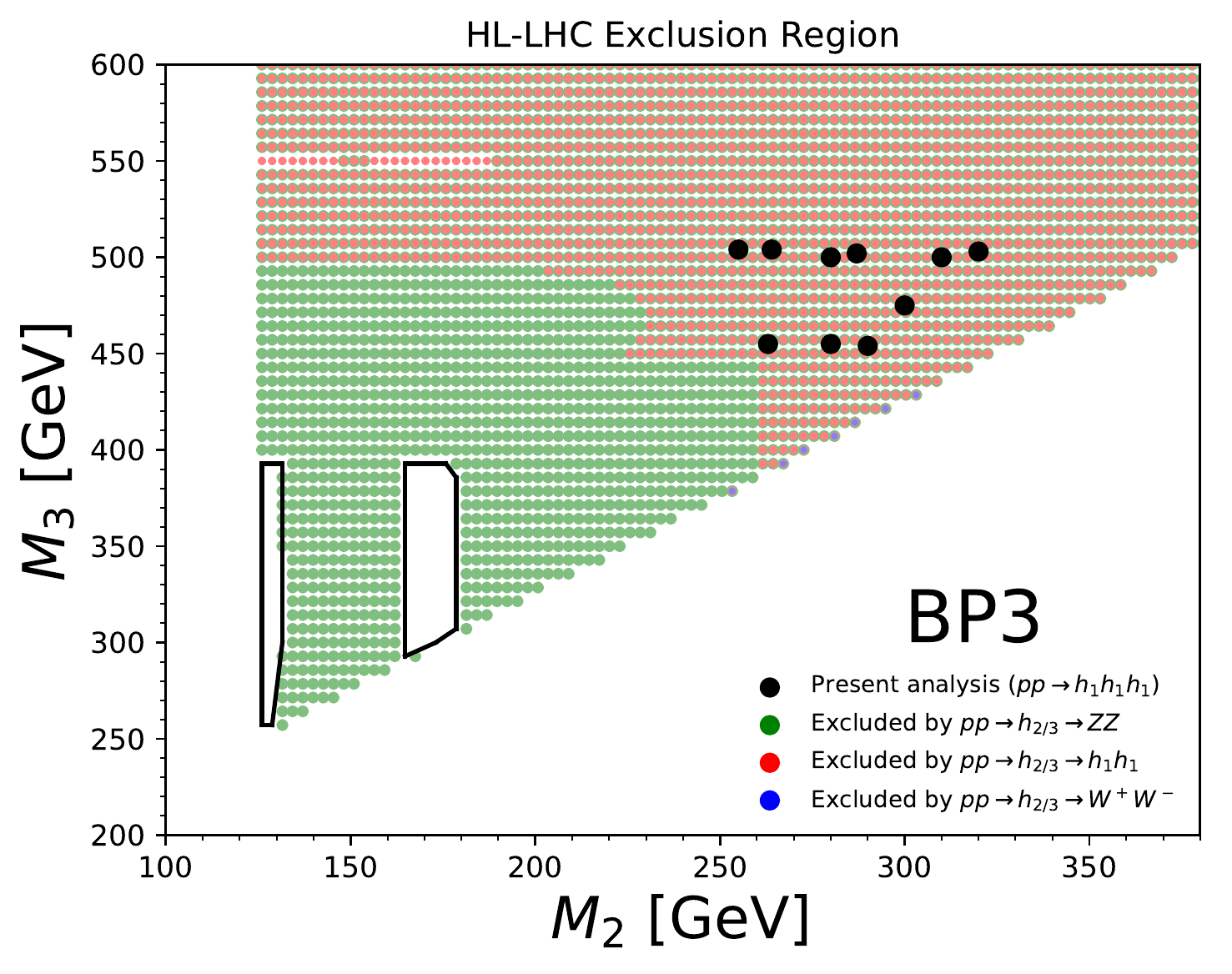}
\caption{\label{fig:const_Andreas} The expected exclusion region for the full integrated luminosity of the HL-LHC, $3000~\rm{fb}^{-1}$, through final states \textit{other} than $p p \rightarrow h_1 h_1 h_1$ as explained in the main text. Points with green circles are expected to be excluded by $ZZ$ final states, with red circles by $h_1 h_1$ and with blue circles by $W^+W^-$. The $W^+W^-$ analysis excludes only very few points on the parameter space and therefore appears infrequently in the figure. The points \textbf{A}--\textbf{I} that we have considered in our analysis of $p p \rightarrow h_1 h_1 h_1$ are shown in black circles overlayed on top of the circles indicating the exclusion. The two cut-out white regions near $M_2 \sim 130$~GeV and $M_2 \sim 170$~GeV will remain viable at the end of the HL-LHC.}
\end{center}
\end{figure}
\end{center}

The decay modes of the $h_2$ and $h_3$ scalars directly into gauge or Higgs boson pairs can also provide signatures for exclusion or discovery in the \textbf{BP3} at the HL-LHC. To investigate these, we have extrapolated various analyses assessing the heavy Higgs boson prospects of the HL-LHC in final states originating from $h_i \rightarrow h_1 h_1$ \cite{Sirunyan:2018two,Aad:2019uzh}, $h_i \rightarrow ZZ$ \cite{Sirunyan:2018qlb,Cepeda:2019klc} and $h_i \rightarrow W^+W^-$ \cite{Aaboud:2017gsl,ATL-PHYS-PUB-2018-022}, for $i=2,3$.  We have combined these with extrapolations of results from 13 TeV where appropriate. For further information, see the detailed analysis presented in Appendix~D of ref.~\cite{Papaefstathiou:2020iag}. The expected exclusion regions for each final state, for an integrated luminosity of $3000~\fb^{-1}$ are displayed in fig.~\ref{fig:const_Andreas}. One can observe that the $ZZ$ final states are by far the most powerful, being capable of excluding almost all of \textbf{BP3} at the HL-LHC. In addition, the $h_1 h_1$ final states will achieve an exclusion of a large fraction of \textbf{BP3}. On the contrary, the $W^+W^-$ final states are foreseen to be rather weak, excluding only a small region of \textbf{BP3}. The significance of the processes in providing exclusion may change in the future if additional decay channels of the gauge or Higgs bosons are considered for each of the processes. Furthermore, detailed experimental studies will be necessary to verify, and potentially improve, our extrapolated observations. 

We note that at the HL-LHC, the effects of the TRSM may also be observed through the reduction of the Higgs boson signal strengths. In \cite{Cepeda:2019klc} a lower limit of $\lb\kappa_{1}^2\rb_\text{min}\,=\,0.933$ was projected for the so-called S1 scenario \cite{uncerthl}, where LHC Run 2 systematic uncertainties were assumed. From table~\ref{tab:BP3}, we see that \textbf{BP3} fulfils this requirement and therefore will not be affected by these measurements. 

\section{Conclusions}\label{sec:summ}

We have examined the triple production of SM-like Higgs bosons, resulting from the asymmetric decay chain $p p \,\rightarrow\, h_3\,\rightarrow\,h_2\,h_1\,\rightarrow\,h_1\,h_1\,h_1,$ within an extension of the SM by two real singlet scalar fields, the TRSM. Our study focused on a specific scenario, ``Benchmark Plane 3'' (\textbf{BP3}) of~\cite{Robens:2019kga}, where current experimental and theoretical constraints are satisfied on a large portion of the plane of masses of the $h_2$ and $h_3$ scalars, $(M_2, M_3)$. We have constructed a Monte Carlo-level phenomenological analysis at the LHC, targeting the $6$ $b$-jet final state originating from the decays of the $h_1$ scalars. Our analysis demonstrates that at an integrated luminosity of $300\,\fb^{-1}$, significances of up to $\sim\,5\,\sigma$ can be achieved for some of the chosen benchmark points on \textbf{BP3}. Furthermore, with the full HL-LHC integrated luminosity of $3000\,\fb^{-1}$, all points that we have considered are within discovery reach, with significances reaching up to $\sim\,16\,\sigma$. We have also shown that gauge or Higgs boson pair final states of the heavy scalars $h_2$ and $h_3$ could probe most of the \textbf{BP3}. 

Our results demonstrate that a combination of all of the examined processes of the present article will be essential to discover \textit{and} gain more insight into the origin of scenarios in which the new physics manifests in a similar manner to \textbf{BP3}. In particular, measurements of the masses of the scalars, the scalar couplings as well as the mixing angles through either single scalar production ($pp \rightarrow h_i$), or multi-scalar production such as the $pp \rightarrow h_1 h_1 h_1$ process of the present article, will allow measurement of the model parameters and reconstruction of the Lagrangian. This will enable model discrimination and a deeper understanding of the r\^ole that such new scalars play in Nature, in case they are discovered. Finally, we emphasise the fact that our analysis indicates that the triple Higgs boson final state, thought to be completely hopeless in the past, should be actively pursued at the LHC through concrete experimental analyses by the ATLAS and CMS collaborations. 

\section*{Acknowledgements}

GTX acknowledges support from the  Deutsche
Forschungsgemeinschaft (DFG, German Research Foundation) under grant  396021762 - TRR 257. AP is supported by the UK's Royal Society. We would like to thank Glen Cowan and Kyle Cranmer for brief discussions on the statistical methods employed. 

\appendix
\section{Scalar Quartic Self-couplings}
We define the quartic scalar self-couplings via
\begin{equation}
V\,\supset\,\sum_{i,j,k,l}\,\lam_{ijkl}\,h_i h_j h_k h_l \;,
\end{equation}
with $i,j,k,l\,=\,1,2,3$. We then have
\begin{eqnarray}
\lambda_{aaaa}&=&\frac{1}{8}\sum_{i,j,k=1}^3\,\frac{M_k^2}{v_i v_j}\,R_{ki}\,R_{kj}\,R^2_{ai}\,R^2_{aj} \;,
\end{eqnarray}

\begin{eqnarray}
\lambda_{aaab}&=&\frac{1}{2}\sum_{i,j,k=1}^3\,\frac{M_k^2}{v_i v_j}\,R_{ki}\,R_{kj}\,R^2_{ai}\,R_{aj}\,R_{bj}\;,
\end{eqnarray}

\begin{eqnarray}
\lambda_{aabc}&=&\frac{1}{2}\,\sum_{i,j,k=1}^3\,\frac{M_k^2}{v_i v_j}R_{ki}R_{kj} R_{ai} R_{cj}\,\lb R_{ai} R_{bj}+2\,R_{bi}\,R_{aj}\rb\;,
\end{eqnarray}

\begin{eqnarray}
\lambda_{aabb}&=&\frac{1}{4}\,\sum_{i,j,k=1}^3\,\frac{M_k^2}{v_i v_j}\,R_{ki} R_{kj} R_{ai}R_{bj}\,\lb R_{ai} R_{bj}+2\,R_{aj} R_{bi}\rb \;,
\end{eqnarray}

for $a \neq b \neq c$.

\section{Total Widths and Branching Ratios}\label{app:vals}

In table~\ref{tab:widths}, we list the total widths as well as decay branching ratios between the physical scalars of the TRSM, for the benchmark points listed in table~\ref{tab:XS_signal}. The total widths have been calculated according to eq.~(\ref{eq:totwidth}), with SM-like widths taken from \cite{Heinemeyer:2013tqa}. Note that the effective branching ratios might vary slightly, as they correspond to $\text{BR}_\text{eff}\,=\,\Gamma^\text{MG5}_{x\,\rightarrow\,y\,z}/\Gamma_x$, where $\Gamma^\text{MG5}_{x\,\rightarrow\,y\,z}$ is the respective partial decay width as calculated by \texttt{MadGraph5\_aMC@NLO}, while $\Gamma_x$ corresponds to the total decay width, which we here treat as an input parameter. For the benchmark points considered here, we however found that deviations are on the sub-percent level. 
\begin{center}
\begin{table}[h!]
\begin{center}
\begin{tabular*}{\textwidth}{@{\extracolsep{\fill}}ccccccc@{}}
Label & $(M_2, M_3)$ &$\Gamma_2$&$\Gamma_3$&$\text{BR}_{2\,\rightarrow\,1\,1}$&$\text{BR}_{3\,\rightarrow\,1\,1}$&$\text{BR}_{3\,\rightarrow\,1\,2}$\\
&  &[\GeV]&[\GeV]&[\GeV]&&\\ \hline
\textbf{A} &$(255, 504)$&0.086&11&0.55&0.16&0.49\\
\textbf{B} &$(263, 455)$&0.12&7.6&0.64&0.17&0.47\\
\textbf{C} &$(287, 502)$&0.21&11&0.70&0.16&0.47\\
\textbf{D} &$(290, 454)$&0.22&7.0&0.70&0.19&0.42\\
\textbf{E} &$(320, 503)$&0.32&10&0.71&0.18&0.45\\
\textbf{F} &$(264, 504)$&0.13&11&0.64&0.16&0.48\\
\textbf{G} &$(280, 455)$&0.18&7.4&0.69&0.18&0.44\\
\textbf{H} &$(300, 475)$&0.25&8.4&0.70&0.18&0.43\\
\textbf{I} &$(310,500)$&0.29&10&0.71&0.17&0.45\\
\textbf{J} &$(280, 500)$&0.18&10.6&0.69&0.16&0.47\\
\end{tabular*}
\caption{\label{tab:widths} The total widths and new scalar branching ratios for the parameter points considered in the analysis. For the SM-like $h_1$, we have $M_1\,=\,125\,\GeV$ and $\Gamma_1\,=\,3.8\,\MeV$ for all points considered. The other input parameters are specified in table~\ref{tab:BP3}. The on-shell channel $h_3\,\rightarrow\,h_2\,h_2$ is kinematically forbidden for all points considered here.}
\end{center}
\end{table}
\end{center}
\section{Combinatorics for Scalar Reconstruction}\label{app:combi}
Here we briefly elaborate further on the scalar reconstruction based on the different arrangements of the 6 $b$-jets with the highest transverse momentum in each event. As discussed in section~\ref{sec:ana}, the aim is to determine the combination of two and three pairs of $b$-jets which minimise the sum

\begin{eqnarray}
\chi^{2,(6)}+\chi^{2, (4)}\;.
\end{eqnarray}

where $\chi^{2,(6)}$  and $\chi^{2, (4)}$ have been introduced in eqs.~(\ref{eq:chi4}) and (\ref{eq:chi6}), respectively.\\

One important aspect of the minimisation is that the set $I$ that defines $\chi^{2, (4)}$ should be a subset of the arrangement $J$ which allows to determine $\chi^{2, (6)}$.\\

Here we achieve our target by using the following procedure

\begin{itemize}
    \item Firstly, we determine all the possible
    combinations of 4 $b$-jets and calculate the corresponding $\chi^{2,(4)}$ for each arrangement of two pairs. We select the configuration $I^{A}_{\rm min}$ with the minimum value of $\chi^{2,(4)A}_{\rm min}$. Notice that, once the arrangement  $I^{A}_{\rm min}$ has been established, there exists only one additional pair of $b$-jets, which allows to complete the configuration of 3 pairs $J^{A}_{\rm min}$, and calculate the corresponding $\chi^{2,(6)}$, denoted as $\chi^{2,(6)A}_{\rm min}$. Then we can compute the sum 
    
    \begin{eqnarray}
    \label{eq:SA}
    S^{A}=\chi^{2,(4)A}_{\rm min} + \chi^{2,(6)A}_{\rm min}\;.
    \end{eqnarray}
    
    \item Subsequently, we obtain all the possible pairings for the full set of 6 $b$-jets and for each one of them we calculate the corresponding $\chi^{2,(6)}$. Out of all the possible configurations we select the combination $J^{B}_{\rm min}$ with the smallest value for $\chi^{2,(6)}$. We label this as $\chi^{2,(6)B}_{\rm min}$. Out of the three pairings that define $J^{B}_{\rm min}$, we can construct 3 possible configurations with two pairs of $b$-jets. We select the one with the minimal $\chi^{2,(4)B}_{\rm min}$ and then we can determine the sum
    
    \begin{eqnarray}
    \label{eq:SB}
    S^{B}=\chi^{2,(4)B}_{\rm min} + \chi^{2,(6)B}_{\rm min}\;.
    \end{eqnarray}
    
    \item Finally, we select the pairs $\{\chi^{2,(4)A}_{\rm min}, \chi^{2,(6)A}_{\rm min}\}$, $\{\chi^{2,(4)B}_{\rm min}, \chi^{2,(6)B}_{\rm min}\}$ with the minimal sum in eqs.~(\ref{eq:SA}) and (\ref{eq:SB}). Thus, if $S^{A}<S^{B}$ then the permutations that enter in the analysis of section~\ref{sec:ana} will correspond to $I_{\rm min}=I^A_{\rm min}$ and $J_{\rm min}=J^A_{\rm min}$ and vice versa.
    
\end{itemize}

Note that this procedure assumes that the $h_1$ bosons are produced on-shell. As discussed in section~\ref{sec:ana}, if $M_2$ and $M_3$ are such that $h_3$ can be produced on-shell through the process $h_3\rightarrow h_1 h_2$ and subsequent $h_2\rightarrow h_1 h_1$, then the configurations $I_{\rm min}$ and $J_{\rm min}$ will ideally correspond to $h_2$ and $h_3$, respectively. Finally, we would like to stress that our optimisation does not assume any a priori values for the masses of the $h_2$ and $h_3$ scalars, i.e. $M_2$ and $M_3$. 

\bibliography{two_scalars}

\end{document}